\documentclass[%
superscriptaddress,
twocolumn,
showpacs,preprintnumbers,
 amsmath,amssymb,
prl,
]{revtex4-1}

\usepackage{graphicx}
\usepackage{color}
\usepackage{dcolumn}
\usepackage{bm}
\usepackage{hyperref}
\usepackage{CJK}

\begin{document}

\title{Thermal Stability and Electrical Control of Magnetization of Heusler/Oxide Interface and Non-collinear Spin Transport of Its Junction}


\author{Zhaoqiang Bai}%
\affiliation {Department of Physics, 2 Science drive 3, National University of Singapore, Singapore 117542, Singapore}%
\affiliation {Data Storage Institute, Agency for Science Technology and Research, 5 Engineering Drive 1, Singapore 117608, Singapore}%

\author{Lei Shen}%
\email{shenlei@nus.edu.sg}%
\affiliation {Department of Physics, 2 Science drive 3, National University of Singapore, Singapore 117542, Singapore}%

\author{Yongqing Cai}
\affiliation {Institute of High Performance Computing, Agency for Science Technology and Research, Fusionopolis, Singapore 138632, Singapore}%

\author{Qingyun Wu}
\affiliation {Department of Physics, 2 Science drive 3, National University of Singapore, Singapore 117542, Singapore}%

\author{Minggang Zeng}
\affiliation {Department of Physics, 2 Science drive 3, National University of Singapore, Singapore 117542, Singapore}%

\author{Guchang Han}%
\affiliation {Data Storage Institute, Agency for Science Technology and Research, 5 Engineering Drive 1, Singapore 117608, Singapore}%

\author{Yuan Ping Feng}%
\email{phyfyp@nus.edu.sg}%
\affiliation {Department of Physics, 2 Science drive 3, National University of Singapore, Singapore 117542, Singapore}%

\date{\today}

\begin{abstract}
 Towards next-generation spintronics devices, such as computer memories and logic chips, it is necessary to satisfy high thermal stability, low-power consumption and high spin-polarization simultaneously. Here, from first-principles, we investigate thermal stability (both structure and magnetization) and the electric field control of magnetic anisotropy on Co$_2$FeAl (CFA)/MgO. A phase diagram of structural thermal stability of the CFA/MgO interface is illustrated. An interfacial perpendicular-anisotropy, coming from the Fe-O orbital hybridization, provides high magnetic thermal stability and a low stray field. We find an electric-field-induced giant modification of such perpendicular-anisotropy via a great magnetoelectric effect (the anisotropy energy coefficient $\beta$$\approx$10$^{-7}$~erg/V~cm). Our spin electronic-structure and \emph{non-collinear} transport calculations indicate high spin-polarized interfacial states and good magnetoresistance properties of CFA/MgO/CFA perpendicular magnetic tunnel junctions.

\end{abstract}

\maketitle


\begin{figure*}
\centering
\includegraphics[width=0.8\textwidth]{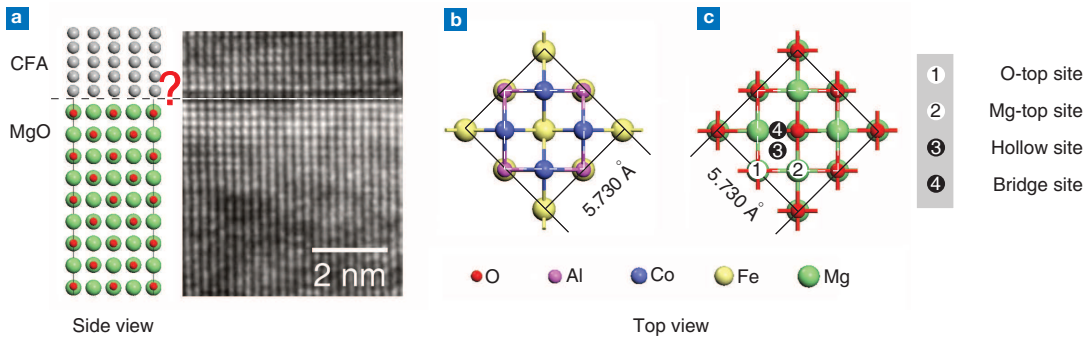}\\
\caption{(Color online) (a) Schematic structure and high-resolution cross-section transmission electron microscopy images of Co$_2$FeAl/MgO(001) [ref.25]. (b) and (c) are interfacial structures of CFA and MgO with an indication of the size of unitcell (the square of white dash line). The different contacted sites are labeled by number 1-4.}
\end{figure*}

Perpendicular magnetic tunnel junctions (p-MTJs), are leading spintronics devices for next-generation computer memories (STT-MRAM) and logic chips for their high thermal stability and good magnetoresistance (MR) properties\cite{Ikeda2010NM}. However, traditional hard ferromagnetic (FM) materials, such as L1$_0$-FePt, are not good electrode materials. It is because of the poor lattice matching with the MgO barrier and low spin-polarization, which would sharply reduce the MR ratio. Furthermore, conversional hard ferromagnets are also inclined to exhibit a large damping constant($\alpha$), e.g., $\alpha_{\textrm{(L1$_0$-FePt)}}$$\geq$0.055 due to the strong spin-orbit interaction (SOI), and hence a high switching current density, i.e., high-power consumption. Experimentally, Ikeda, $et~al.$ proposed the perpendicularly magnetized CoFe (a soft ferromagnet) in ultra-thin CoFe/MgO/CoFe MTJs to be effective for achieving high thermal stability and high MR ratio (124\%)\cite{Ikeda2010NM}. However, the damping constant of the CoFe film increases sharply with the decreasing thickness. Recently, an electric-field-assisted magnetic switching is demonstrated in CoFe/MgO/CoFe MTJs, which can efficiently reduce the switching current density\cite{Wang2012NM}. It provides a more energy-efficient route and is compatible with current ubiquitous electrically-controlled semiconductor devices. Actually, electric control of magnetization has been widely studied (both experimentally and theoretically) in ferromagnetic thin films\cite{Duan2006PRL}, FM/metal\cite{Nakamura2009PRL,Tsujikawa2009PRL,Tsujikawa2008PRB}, FM/oxide interfaces, such as FePt/MgO\cite{Weisheit2007Science,Zhu2011APL}, Fe/MgO\cite{Maruyama2009NN,Niranjan2010APL,Nakamura2010PRB} and CoFe/MgO\cite{Wang2012NM,He2011APL}. Although the coefficient of magnetic anisotropy energy, $\beta$, ($\Delta_{\textrm{MCA}}=\beta\times{E^\textrm{F}}$, where $\Delta_{\textrm{MCA}}$ is the change in the magnetic anisotropy and $E^\textrm{F}$ is the electric field.) of FM/oxide is one order of magnitude higher than that of pure Fe surfaces\cite{Duan2006PRL} and FM/metal interfaces\cite{Nakamura2009PRL,Tsujikawa2009PRL,Tsujikawa2008PRB}, but it is still quite low ($\approx$10$^{-8}$~erg/V~cm) for practical low-power applications. That is to say, the electrically-controlled energy-efficiency is quite low now. Therefore, To find a FM/oxide interface with a giant modification of magnetization by an electric field is a matter of great urgency. Meanwhile, it is still absent of FM/oxide/FM junctions to satisfy high thermal stability, low switching current density (i.e., low-power consumption) and good spin transport properties all at the same time.

In this Letter, we predict a great magnetoelectric effect at the interface of Co$_2$FeAl/MgO (001). The anisotropy energy coefficient, $\beta$, is one order of magnitude higher than that of traditional FM/MgO interfaces. Furthermore, high thermal stability (both structure and magnetization) is demonstrated in the FeAl$\mid$O interfacial structure. The great magnetoelectric effect (CFA/MgO) combined with a small Gilbert damping constant (Co$_2$FeAl) and low stray field (perpendicular anisotropy) means a low critical current density in CFA/MgO-based MTJs. On the basis of this interfacial structure, we build a junction (Co$_2$FeAl/MgO/Co$_2$FeAl) which shows a high spin-polarized interfacial states, and our \emph{non-collinear} spin-transport calculations show this p-MTJ has better MR properties (MR ratio and $\Delta$RA) than conventional CoFe/MgO/CoFe and FePt/MgO/FePt MTJs. The high thermal stability, low-power composition, and good MR properties indicate CFA/MgO/CFA MTJs to be one of the best candidates for next-generation low-power/high-performance spintronics devices.


A CFA/MgO(001) interface was modeled as shown in \textbf{Fig.~1a}. The in-plane lattice parameter was fixed at 1/$\sqrt{2}$ of the experimental lattice parameter of bulk CFA (5.730~\AA) with a very small lattice mismatching with MgO ($\sim$3.7\%). The electronic structure calculations of CFA/MgO (001) interfaces were carried out using a first-principles approach[see \textbf{Supplemental Materials}]. The non-equilibrium Green's function formulism was used for the non-collinear transport calculation of CFA/MgO/CFA p-MTJs [see \textbf{Supplemental Materials}].


Recently, a half-metallic Heusler compound (Co$_2$FeAl) is demonstrated the perpendicular magnetic anisotropy (PMA) in CFA/Pt multilayers prepared on MgO (001) substrates\cite{Wang2010APE}. Subsequently, Wen $et~al.$ reported a PMA in the structure of CFA/MgO with a high $K_\textrm{u}$ of 2-3$\times$10$^6$~erg/cm$^3$ \cite{Wen2011APL}. Before any magnetoelectric or transport studies on CFA/MgO structures, the first and important thing is to determine the interfacial thermal stability structurally and magnetically. For example, although the high-resolution transmission electron microscopy (HRTEM) can give a clear atomic structure of the CFA/MgO interface, it is still not clear the distribution of different types of atoms at the interface (see \textbf{Fig.~1a}). Furthermore, it is known that the interfacial geometry strongly affects the spin interfacial states and spin transport properties\cite{Duan2006PRL,Ke2008PRL,Ke2010PRL}, while only a few studies are carried on the the interfacial structure of ternary Heusler alloys with oxides because it is much more complex than that of CoFeB after high-temperature annealing\cite{Hulsen2009PRL,Miura2008PRB,Bai2013PRB}. Therefore, we first study the thermal stability of the interfacial geometry of CFA/MgO. \textbf{Figure~1b and 1c} show that the CFA/MgO interface has four possible types of CFA-terminations, i.e., CoCo, FeAl, FeFe, and AlAl. Furthermore, each of these terminations has four possible positions relative to MgO (001), i.e., O-top, Mg-top, hollow and bridge site. The total energy calculations show that for the CoCo-, FeAl-, and FeFe-terminated CFA, the configuration of the O-top site is lower in energy than any others, whereas for the AlAl-termination, the Mg-top site is the most energetically favorable. Thus, we use the method of $ab~initio$ atomistic thermodynamics to compare the stability among different terminations\cite{Weinert1987PRL,Hulsen2009PRL}. The interface formation energy is calculated by the formula:
\begin{eqnarray}
\gamma(T,p)=\frac{1}{2A}[\Delta G(N_i,T,p)-\Sigma_i\Delta N_i\mu_i(T,p)]
\label{eq:one}.
\end{eqnarray}

where $\Delta{G}$ is the difference in the Gibbs free energy between the heterostructure and appropriate amounts of two bulk materials forming the interface. $\Delta N_i$ ($i$ = Co, Fe or Al) counts the number of atoms, in which the supercell deviates from bulk composition, and $\mu_i$ represents the chemical potential of the element $i$. Since the interface energy is a free-energy difference, it is often a good approximation to calculate $\gamma$ from the differences of total energies obtained from DFT calculations\cite{Hashemifar2005PRL}. For the thermodynamic analysis, it is assumed that the interface is in thermodynamic equilibrium with bulk Heusler compounds. This allows us to eliminate one chemical potential term with the relation: $g_{\textrm{CFA}}=2\mu_{\textrm{Co}}+\mu_{\textrm{Fe}}+\mu_{\textrm{Al}}$, where $g_{\textrm{CFA}}$ represents the Gibbs free energy of Co$_2$FeAl per formula unit. Hereby, only two of the three chemical potential values $\mu_i$ are independent. Under the Co- or Fe-rich condition, Co or Fe is assumed in thermodynamic equilibrium with its bulk solid phase ($\mu_{\textrm{Co(bulk)}}$ and $\mu_{\textrm{Fe(bulk)}}$ ) because the excess Co or Fe atoms may lead clusters or even precipitation. Consequently, the chemical potential $\mu_{\textrm{Co}}$ and $\mu_{\textrm{Fe}}$ may not exceed their chemical potential of bulk phase, $\mu_{\textrm{Co}}\leq\mu_{\textrm{Co(bulk)}}$ and $\mu_{\textrm{Fe}}\leq\mu_{\textrm{Fe(bulk)}}$. The interface phase diagram (\textbf{Fig.~2}) displays the termination with the minimal interface energy for every chemical potential pair ($\mu_{\textrm{Co}}$ and $\mu_{\textrm{Fe}}$). It can be concluded from the phase diagram that two termination configurations of the Heusler compound, i.e., the CoCo- and FeAl-planes sitting on top of oxygen, are thermodynamically stable. In contrast, the other two terminations are not accessible within the limit set by thermodynamic equilibrium with other phases. Further analysis on the phase diagram shows that, in the wide range of $\mu_{\textrm{Co}}$, the FeAl-termination is more favorable than the CoCo-termination because almost all the orange area are contributed by the green area. In experiments, the Co$_2$FeAl film is high-temperature annealed after deposition on a buffer layer in order to obtain the ordered L2$_1$ phase. Therefore, the most thermodynamically stable interface structure after annealing in the experiment should be the one derived from our chemical potential analysis above, i.e., FeAl$\mid$O. Thus, we following focus on the FeAl-terminated Co$_2$FeAl/MgO(001) interface and discuss its magnetocrystalline anisotropy, magnetoelectric effect and spin transport.

\begin{figure}
\centering
\includegraphics[width=0.40\textwidth]{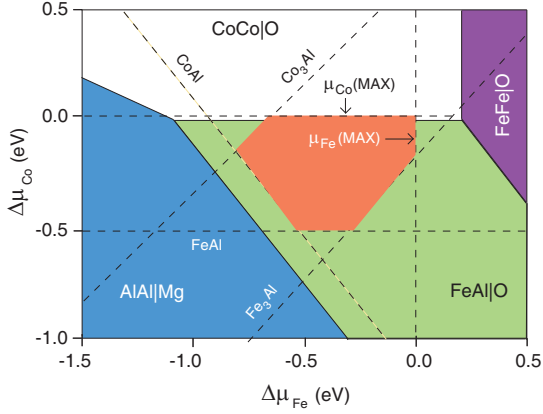}\\
\caption{(Color online) Phase diagram for the epitaxial CFA/MgO (001) interface. The colored regions correspond to different interface terminations being stable under the conditions described by the chemical potentials $\Delta{\mu}_i=\mu_i-g_i$, where $g_i$ ($i=$Co,Fe) is the total energy of one unitcell. The orange hex-polygon indicates the region accessible in thermodynamic equilibrium with the bulk phases of Fe, Co, Fe$_3$Si, Co$_3$Si, FeSi, or CoSi.}
\end{figure}

\begin{figure}
\centering
\includegraphics[width=0.40\textwidth]{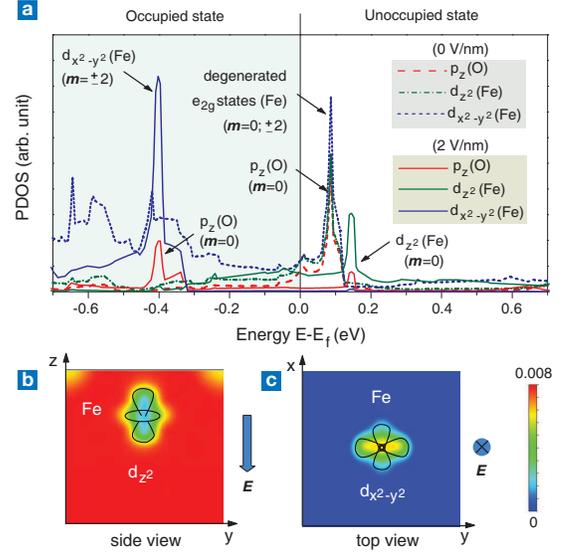}\\
\caption{(Color online) (a) The orbital-resolved PDOSs projected on the interfacial iron and oxygen atoms in absence of electric fields (dash line) and under the electric field of 2~V/nm (solid line). Induced differential charge density, in units of e/\AA$^3$, at the interfacial Fe atom for E=2~V/nm in the (100) plane (b) and (001) plane (c).}
\end{figure}


Besides the structurally thermal stability, the magnetically thermal stability (hard magnetism with PMA) at the interface plays another key role on the device performance because the perpendicular magnetic anisotropy ($K_\textrm{u}$) can beat the thermal perturbation in ambient conditions ($k_\textrm{B}T$, where, $k_\textrm{B}$ and $T$ are Boltzman constant and temperature, respectively). Without PMA, the MR ratio of MTJs in room temperature would be much lower than that in low temperature due to the thermal perturbation on the interfacial magnetization\cite{Miura2011PRB}. Moreover, the perpendicular anisotropy can strongly reduce the stray field, and then the power consumption of devices. For CFA/MgO, our calculations (involving SOI) show that perpendicular magnetocrystalline anisotropy is energetically favored with an anisotropy constant $K_\textrm{u}$ of 0.428~erg/cm$^2$, which is in good agreement with the previous experimental report\cite{Wen2011APL}. The origin of PMA can be understood by the electronic structure around the Fermi level via the second order perturbation consideration\cite{Shen2013AM,Nakamura2009PRL,Tsujikawa2009PRL}:

\begin{eqnarray}
MCA\propto \sum_{\textbf{k}}\sum_{o,u}\frac{|<\textbf{k}_o|L_z|\textbf{k}_u>|^2-|<\textbf{k}_o|L_x|\textbf{k}_u>|^2}{\varepsilon_{\textbf{k}_u}-\varepsilon_{\textbf{k}_o}}
\label{eq:three}.
\end{eqnarray}

where MCA is the magnetocrystalline anisotropy; $\textrm{\textbf{k}}_o$ and $\textrm{\textbf{k}}_u$ represent the occupied and unoccupied states with the wave vector $\textrm{\textbf{k}}$; and $L_z$ and $L_x$ are the angular momentum operators along (001) and (100) direction. According to \textbf{Eq.~2}, the SOI between the occupied and unoccupied states with the same magnetic quantum number (\textit{m}) through the $L_z$ operator enhances $E_{\textrm{MCA}}(\textbf{k})$, while that with different \textit{m} through the $L_x$ operator weakens it. \textbf{Figure 3} shows the projected density of states (PDOS) of CFA/MgO. Only the minority-states are shown for clarity, since the MCA originates mainly from the SOI between the minority spin bands around the Fermi level\cite{Shen2013AM,Tsujikawa2009PRL,Nakamura2009PRL}.
A metal-induced gap state (MIGS) is occurred near the Fermi energy, which is composed of O-$p_z$ and Fe-$e_{2g}$ orbitals ($d_{z^2}$ and $d_{x^2-y^2}$). Thus, on the basis of the above perturbation theory, the spin-orbit coupling between O-$p_z$ and Fe-$d_{z^2}$  (with the same quantum number 0) through the $L_z$ operator (001) contributes positively to the PMA in sense that the matrix element $<p_z|L_z|d_{z^2}>$ is non-vanishing, and hence explains the perpendicular (001) magnetic anisotropy. Note that the contribution to MCA of $<p_z|L_z|d_{x^2-y^2}>$ and $<p_z|L_x|d_{x^2-y^2}>$ are zero.

\begin{figure}
\centering
\includegraphics[width=0.45\textwidth]{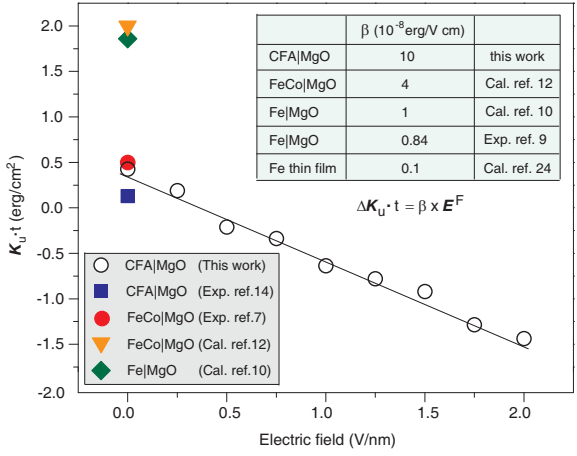}\\
\caption{(Color online) The magnetic anisotropy energy density ($t$=1~nm) of CFA/MgO as a function of the external electric field. Some experimental and calculated $K_\textrm{u}\cdot{t}$ of CFA/MgO, CoFe/MgO and Fe/MgO (without $E^\textrm{F}$) are shown for comparison. The inset table shows the anisotropy coefficient ($\beta$) of some FM/MgO structures (with $E^\textrm{F}$).}
\end{figure}

Recently, a manner of efficiently manipulating various magnetic properties by the application of an external electric field, namely a magnetoelectric effect, has been successfully demonstrated theoretically and experimentally\cite{Duan2006PRL,Ikeda2010NM,Maruyama2009NN,Tsujikawa2009PRL,Nakamura2009PRL}. As mentioned in the introduction, electrical control of magnetization is not only more power saving, but also compatible with current ubiquitous electrically-controlled semiconductor devices for the large-scale integrated circuit design. Thus, we next investigate the electric-field-induced modification of MCA of CFA/MgO. In \textbf{Fig.~4}, our calculated anisotropy energy density of CFA($t=$1~nm)/MgO is in good agreement with previous experimental results\cite{Wen2011APL} and in the same order of magnitude with calculated CoFe/MgO\cite{Ikeda2010NM} or Fe/MgO\cite{Maruyama2009NN}. Our calculated values of the MCA density show a almost linear dependence on the external electric field. The interfacial MCA coefficient $\beta$, defined as $\Delta{K_\textrm{u}}\cdot{t}=\beta\times{E^\textrm{F}}$, where $\Delta{\textrm{K}}$ is the change in the MCA energy, a key index for the strength of the MCA modification by an electric field. Surprisedly, the value of $\beta$ of CFA/MgO is one order of magnitude larger than that of CoFe/MgO\cite{Ikeda2010NM,He2011APL} or Fe/MgO\cite{Maruyama2009NN,Niranjan2010APL} and almost two order of magnitude larger than that of Fe thin film or Fe/metal~(001)\cite{Duan2008PRL,Tsujikawa2009PRL,Nakamura2009PRL}. Such giant modification of MCA by the electric field in CFA/MgO indicates a more energy-efficient route from the electric control of magnetization for low-power device applications.

It is found that the above great ME effect derives from the electrical-field-induced redistribution of spin electrons between difference orbitals around the Fermi level. In \textbf{Fig.~3a}, the originally degenerated high-lying Fe-$e_{2g}$ states are split by the external electrical field with an exchange splitting energy of 0.53~eV. Such splitting pushes the Fe-$d_{x^2-y^2}$ orbital below the Fermi level of 0.4~eV (thus fully occupied). Meanwhile, the O-$p_z$ orbital follows up the shift of Fe-$d_{x^2-y^2}$ orbital and always strongly couples with Fe-$d_{x^2-y^2}$ independent on the external electric field. The Fe-$d_{z^2}$ is shifted above the Fermi level by the electric field, indicating a charge transfer from Fe-$d_{z^2}$ to Fe-$d_{x^2-y^2}$ and O-$p_z$. Consequently, the mentioned origin of PMA ($<p_z|L_z|d_{z^2}>$) is completely destroyed due to the decoupling between O-$p_z$ and Fe-$d_{z^2}$ [see \textbf{Fig.~3a}]. A further quantitative illustration of such $d$-electron redistribution is shown in \textbf{Fig.~3b} and \textbf{3c} by plotting the induced charge density (differential) on interfacial Fe atoms, $\Delta{\rho_{\textrm{Fe}}}=\rho_{\textrm{Fe}}(2.0)-\rho_{\textrm{Fe}}(0)$, within the energy window $E_\textrm{f}-0.5~\textrm{eV}<E<E_\textrm{f}+0.2~\textrm{eV}$, where the specified charge transfer takes place. \textbf{Figure~3b} and \textbf{3c} clearly show a reduced occupation of the Fe-$d_{z^2}$ orbitals and an enhanced occupation of Fe-$d_{x^2-y^2}$ orbitals.

\begin{figure}
\centering
\includegraphics[width=0.4\textwidth]{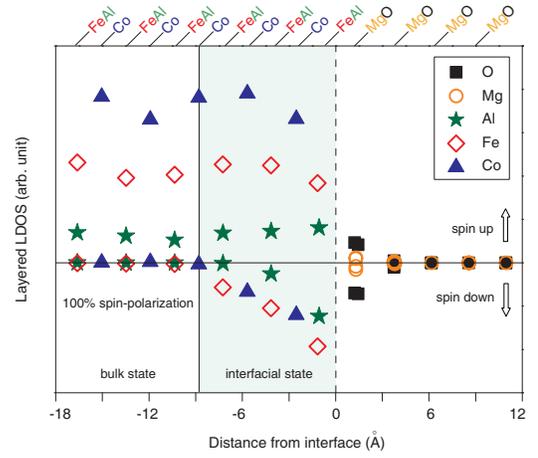}\\
\caption{(Color online) The spin-resolved LDOSs at the Fermi energy projected on to each atomic sphere as a function of the distance from the interface for Co$_2$FeAl/MgO/Co$_2$FeAl (001). The positive (negative) sign of \textit{y} axis indicates spin up (down).}
\end{figure}

We note that in the experiment the L1$_2$-FePt/MgO interfacial structure also shows good interfacial properties, such as large PMA ($K_u$$\sim$10$^6$-10$^7$ erg/cm$^3$) and giant anisotropy energy coefficient ($\beta$$\sim$10$^{-7}$~erg/V~cm)\cite{Weisheit2007Science}. However, FePt/MgO/FePt p-MTJs show worse MR properties, neither TMR ratio nor $\Delta$MR, because of low spin-polarized interface states\cite{Wen2011APL}. This raises the question of whether CFA/MgO (with good interfacial properties) has good transport (i.e., MR) properties for practical device applications? With this question, we next investigate the spin-polarization of CFA/MgO/CFA p-MTJs, including interfacial spin polarization and non-collinear spin transport properties.

It is well known that the spin-dependent tunneling conductance of the MgO-based MTJs is determined by two crucial factors. One is the symmetry matching between the transmission Bloch states in FM electrodes and evanescent states in the MgO barrier. The other is the spin-polarization of the FM/MgO interfacial states\cite{Bai2013PRB,Ke2008PRL,Ke2010PRL}. The Bloch states with $\Delta$1 symmetry suffer from the smallest decay in MgO\cite{Bai2013PRB}. Therefore, the spin-filter efficiency can be tremendously enhanced if the FM electrodes have $\Delta$1-band half-metallicity. It has been reported that, for both bulk CFA\cite{Wang2010PRB} and CoFe\cite{Bai2013PRB} electrodes, only majority bands which cut the Fermi level possess the $\Delta$1 feature and fulfill the symmetry-matching criterion. In contrast, $\Delta$1-band of L1$_\textrm{2}$-FePt predominantly contributes to the tunneling conductance for both the majority and the minority-spin channels.\cite{Taniguchi2008IEEE} \textbf{Figure~5} illustrates the spin-resolved layered LDOSs at the Fermi energy projected on to each atomic sphere as a function of the distance from the interface for Co$_2$FeAl/MgO/Co$_2$FeAl. As can be seen, low spin-polarized interfacial states only appear on the first atomic layer. High spin-polarization is quickly resumed in next a few atomic layers. For example, the spin-polarization is recovered to 75\% in the second atomic layer and it becomes 100\% (i.e., half-metal) of 0.8~nm thickness.

The non-equilibrium Green's function-based collinear spin transport calculations can provide a good description of MR properties of conventional MTJs and good explanation on many experimental observed phenomenon\cite{Ke2008PRL,Ke2010PRL,Bai2013PRB}. However, the non-collinear transport method must be considered into the ``perpendicular" MTJs. We, for the first time, use the state-of-art non-collinear coherent transport method [see \textbf{Supplemental Materials}] to study the spin transport of CFA/MgO/CFA (001) p-MTJs. For comparison, the widely-used CoFe/MgO/CoFe (001) and FePt/MgO/FePt (001) junctions are also included in this discussion. The calculated conductance values are listed in \textrm{Table~1}. As can be seen, ...... Thus, CFA should be, at least in principle, an eligible candidate to substitute the conventional FM materials as the electrode material of the MgO-based MTJs from the perspective of enhanced MR performance.

\begin{table}
\caption{\label{tab:table1}The calculated conductance (Siemens) of the parallel majority ($G^{maj}_{\leftarrow\leftarrow}$), parallel minority ($G^{min}_{\rightarrow\rightarrow}$) and antiparallel ($G^{maj/min}_{\leftarrow\rightarrow}$) channels of the CFA/MgO/CFA, CoFe/MgO/CoFe, and FePt/MgO/FePt perpendicular magnetic tunnel junctions. The MgO spacer length is 17.1~\AA~for all three structures.}
\centering
\begin{tabular}{l|cccc}
\hline
\hline
Structure  & $G^{maj}_{\leftarrow\leftarrow}$ & $G^{min}_{\rightarrow\rightarrow}$ & $G^{maj/min}_{\leftarrow\rightarrow}$ \\
\hline
CFA/MgO/CFA  & 1.38$\times$10$^{-9}$ & 9.732$\times$10$^{-20}$ & 3.37$\times$10$^{-14}$ \\
\hline
CoFe/MgO/CoFe  & 2.72$\times$10$^{-10}$ & 2.73$\times$10$^{-13}$ & 6.45$\times$10$^{-12}$ \\
FePt/MgO/FePt   & 3.23$\times$10$^{-11}$ & 9.99$\times$10$^{-13}$ & 5.54$\times$10$^{-12}$ \\
\hline
\hline
\end{tabular}
\end{table}

In conclusion, we provide a phase diagram of interfacial thermal stability of Co$_2$FeAl/MgO as a guide for experimentalists. A perpendicular magnetic anisotropy is demonstrated in CFA/MgO, which not only provides a magnetically thermal stability, but also reduces the stray field. An electric-field-induced giant modification of the PMA (a great magnetoelectric effect) combined with a low damping constant and small stray field indicates a potential low-power and high compatible spintronics device application of CFA/MgO structures. We, thus, model a CFA/MgO/CFA (001) magnetic tunnel junction and find a high spin-polarized interfacial states and quite good magnetoresistance properties (large MR ratio and $\Delta$MR) from non-collinear spin transport calculations. It shows that Heusler (CFA) based perpendicular magnetic tunnel junctions hold the promise for next-generation spintronics applications, such as non-volatile computer memories and logic chips\cite{Bai2012SPIN}. We should point out that this prediction is just a proof-of-principle study, further experimental studies are needed.

The Authors thank Kurt Stokbro and Anders Blom for privately providing the state-of-art non-collinear transport calculation package (ATK v13.8). Bai thanks Zhenchao Wen and Zhen Liu for their helpful discussion on the growth and characterization of CFA/MgO (001). This work is partially supported by A*STAR (Singapore) Research Funding (Grant No. 092-156-0121).

%



\end{document}